\documentclass{article}
\usepackage{spconf,amsmath,graphicx}
\usepackage{array}
\usepackage{multirow}
\usepackage{graphicx}
\usepackage{enumerate}
\usepackage{setspace}
\usepackage[skip=2pt,font=small]{caption}
\usepackage[shortlabels]{enumitem}
\usepackage{url}

\usepackage{hyperref}

\newcommand*{\Scale}[2][4]{\scalebox{#1}{$#2$}} 

\title{Completed Local Derivative Pattern for Rotation Invariant Texture Classification}
\name{Yuting Hu, Zhiling Long, and Ghassan AlRegib}
\address{Multimedia \& Sensors Lab (MSL)\\
Center for Signal and Information Processing (CSIP)\\
School of Electrical and Computer Engineering\\
Georgia Institute of Technology, Atlanta, GA 30332-0250, USA\\
\{huyuting, zhiling.long, alregib\}@gatech.edu}
\begin{document}

\onecolumn 

\begin{description}[labelindent=1cm,leftmargin=4cm,style=multiline]

\item[\textbf{Citation}]{Y. Hu, Z. Long, and G. AlRegib, ``Completed Local Derivative Pattern for Rotation Invariant Texture Classification,'' IEEE International Conference on Image Processing (ICIP 2016), pp. 3548-3552, 2016.}

\item[\textbf{DOI}]{\url{https://doi.org/10.1109/ICIP.2016.7533020}}

\item[\textbf{Review}]{Date of publication: 19 August 2016}

\item[\textbf{Codes}]{\url{https://ghassanalregibdotcom.files.wordpress.com/2016/10/yuting_icip2016_code.zip}}

\item[\textbf{Bib}] {@inproceedings{hu2016completed,\\
  title={Completed local derivative pattern for rotation invariant texture classification},\\
  author={Hu, Yuting and Long, Zhiling and AlRegib, Ghassan},\\
  booktitle={Image Processing (ICIP), 2016 IEEE International Conference on},\\
  pages={3548--3552},\\
  year={2016},\\
  organization={IEEE}
}
}


\item[\textbf{Copyright}]{\textcopyright 2018 IEEE. Personal use of this material is permitted. Permission from IEEE must be obtained for all other uses, in any current or future media, including reprinting/republishing this material for advertising or promotional purposes,
creating new collective works, for resale or redistribution to servers or lists, or reuse of any copyrighted component
of this work in other works. }

\item[\textbf{Contact}]{\href{mailto:huyuting@gatech.edu}{huyuting@gatech.edu}  OR \href{mailto:zhiling.long@ece.gatech.edu}{zhiling.long@ece.gatech.edu} OR \href{mailto:alregib@gatech.edu}{alregib@gatech.edu}\\
    \url{http://ghassanalregib.com/} \\ }

\end{description}

\thispagestyle{empty}
\newpage
\clearpage
\setcounter{page}{1}

\twocolumn

%
\maketitle
\begin{abstract}
 In this paper, we propose a new texture descriptor, completed local derivative pattern (CLDP). In contrast to completed local binary pattern (CLBP), which involves only local differences at each scale, CLDP encodes the directional variation of the local differences of two scales as a complementary component to local patterns in CLBP. The new component in CLDP, with regarded as the directional derivative pattern, reflects the directional smoothness of local textures without increasing computation complexity. Experimental results on the Outex database show that CLDP, as a uni-scale pattern, outperforms uni-scale state-of-the-art texture descriptors on texture classification and has comparable performance with multi-scale texture descriptors.
\end{abstract}
\section{Introduction}
\label{sec:intro}

\begin{figure*}[t]
    \begin{minipage}[b]{1.0\linewidth}
      \centering
      \centerline{\includegraphics[width=15.5cm]{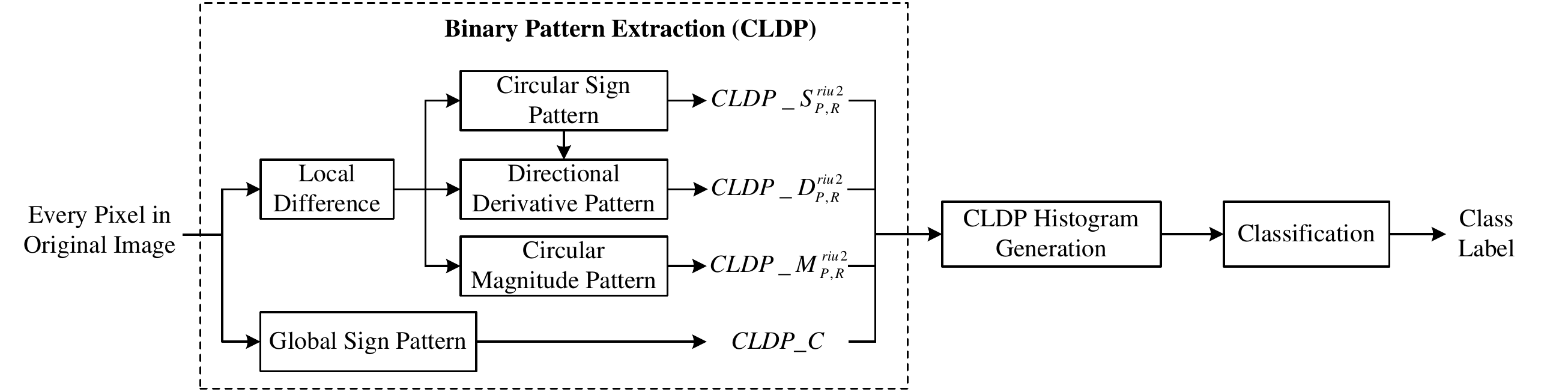}}
    \end{minipage}
    \caption{The block diagram of the proposed CLDP-based texture classification method}
    \label{fig:algorithm framework}
\end{figure*}

Texture descriptors are widely used in applications of computer vision such as image retrieval~\cite{cai2014scalable} and image matching~\cite{rublee2011orb}. One typical example is the local binary pattern (LBP) proposed by Ojala et al.~\cite{ojala1996comparative}, which describes the sign information of local differences between each pixel and its neighbors. By encoding the sign information in a circular manner into binary codes, LBP can achieve high efficiency on texture representation.

Because of LBP's great success in face recognition and texture classification, several methods derived from LBP have been introduced in recent years. Guo et al.~\cite{guo2010completed} proposed CLBP, which extracts not only local sign and magnitude information, but also global intensity information. Without uniform patterns in LBP, Liao et al.~\cite{liao2009dominant} proposed the dominant local binary pattern (DLBP), which computes the occurrence frequencies of all rotation invariant LBP patterns and defines the most frequent ones as dominant patterns. Moreover, to enhance classification performance and robustness, LBP-based algorithms~\cite{guo2010completed,zhao2012completed,guo2012local,lin2015multi,qi2013multi} involving multi-scale patterns have been proposed. However, almost all of these LBP-based methods ignore local derivatives, which contain complementary discriminative information~\cite{zhang2010local}. To address this drawback, Zhang et al.~\cite{zhang2010local} proposed the local derivative pattern (LDP) to encode the local derivative information of various directions. Although LDP shows good performance on face recognition, it cannot ensure rotation invariance in texture classification without a circular coding strategy. To accomplish rotation invariance in local derivative patterns, Guo et al.~\cite{guo2012local} proposed the local directional derivative pattern (LDDP). However, the performance of LDDP is not comparable to that of CLBP in texture classification because of the lack of magnitude information.

In this paper, we propose the framework of a new texture descriptor, CLDP, to represent local texture features. In contrast to CLBP, which encodes local binary patterns in each scale separately, CLDP adds a new component, the directional derivative pattern, which involves the patterns of two neighboring scales in the same direction to calculate the corresponding directional cross-scale correlation. The directional derivative pattern in CLDP characterizes local texture smoothness along each direction. Therefore, we utilize four types of patterns in CLDP to represent the sign, magnitude, and local directional derivative information in local differences and the intensity values of center pixels, respectively. We combine these four patterns using histogram-based manners and generate feature vectors for texture classification. To verify the performance of CLDP, we compare it with state-of-the-art uni- and multi-scale texture descriptors on the Outex database~\cite{ojala2002outex}. The experimental results show that CLDP outperforms its CLBP counterpart in texture classification accuracy without adding computational complexity. Furthermore, the proposed CLDP method has better classification performance than the uni-scale descriptors and is comparable to the multi-scale texture descriptors on the classification accuracy as well.


The rest of the paper is organized as follows. Section~\ref{sec:cldp} introduces the proposed texture descriptor CLDP and its application in texture classification. Section~\ref{sec:experimental results} presents experimental results on the Outex database. Section~\ref{sec:conclusions} makes a conclusion.
\vspace{-0.1in}
\section{Proposed Method}
\label{sec:cldp}
The block diagram of the proposed method is shown in Fig.~\ref{fig:algorithm framework}. In following subsections, we are going to introduce each block in detail. Our notations here follow those of~\cite{guo2010completed}.
\vspace{-0.15in}
\subsection{Binary Pattern Extraction}
\label{ssec:binary pattern extraction}

\subsubsection{Local Difference}
\label{sssec:local difference}
\begin{figure}[t]
    \centering
    \begin{minipage}[!htbp]{1\linewidth}
      \centering
      \centerline{\includegraphics[width=5cm]{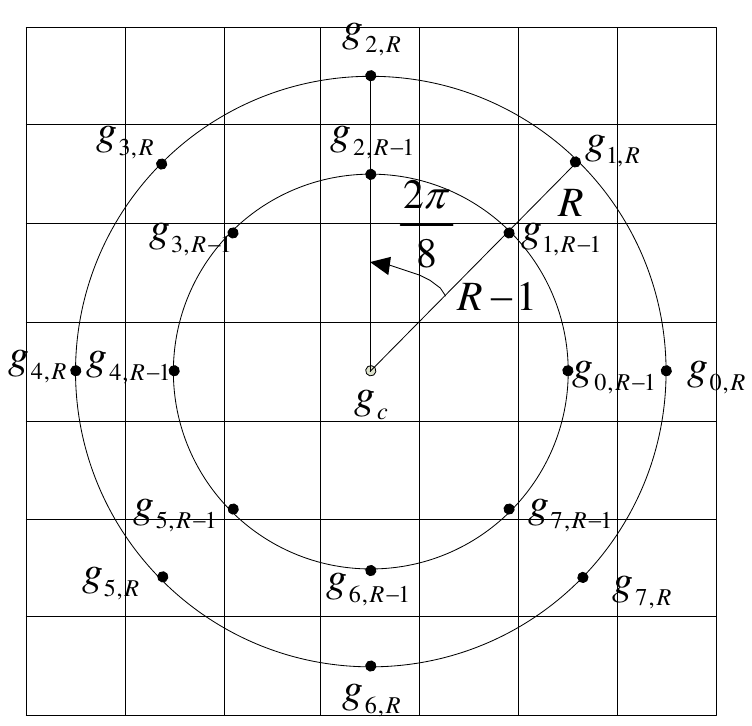}}
    \end{minipage}
\caption{The sampling scheme for $P=8$ with radii $R$ and $R-1$}
\label{fig:sampleblock}
\end{figure}

The local difference has been widely used in texture representation because of its robustness to illumination changes. As Fig.~\ref{fig:sampleblock} shows, each pixel $g_c$ in the texture image corresponds to $P$ neighbors, which are evenly distributed on a circle with radius $R$. We denote these neighbors as $g_{p,R}, p=0,1,\cdots,P-1$. If $g_{p,R}$ does not have integer coordinates, its intensity value can be estimated by bilinear interpolation. On the basis of $g_{p,R}$, we define the local difference, denoted $d_{p,R}$, as $d_{p,R}=g_{p,R}-g_c$.

\vspace{-0.15in}

\subsubsection{Circular Sign Pattern}
\label{sssec:circular sign pattern}
According to local difference, we define the sign component of CLDP as $s_{p,R}=\left\{\begin{aligned}&1,\ d_{p,R}\geq0\\&0,\ d_{p,R}<0
\end{aligned}\right.$~\cite{guo2010completed}. For simplification, we denote such a thresholding function as $s_{p,R}=t\left(d_{p,R},0\right)$, where 0 represents the threshold. By applying function $t\left(d_{p,R},0\right)$ on all directions, we obtain $P$ binary bits and encode them in a counter-clockwise manner, denoted $CLDP\_S_{P,R}$. To guarantee the rotation invariance in texture classification, we group binary codes with the same circularly shifted format into one pattern and denote the new pattern as $CLDP\_S_{P,R}^{ri}$, where ``$ri$'' represents rotation invariance. For example, if $P=8$, we can reduce the total number of patterns from $2^8=256$ to $36$~\cite{ojala2002multiresolution}. In addition, to further reduce the number of patterns, researchers commonly involve another criterion, the uniform pattern, denoted as ``$u2$'', in which the frequency of bitwise transitions from $0$ to $1$ or $1$ to $0$ is less than two.
Therefore, the uniform patterns of $CLDP\_S_{P,R}^{ri}$, denoted $CLDP\_S_{P,R}^{riu2}$, can be calculated as follows:
\begin{equation}
\label{eq:CLDP_S_riu2}
\Scale[0.95]{
CLDP\_S_{P,R}^{riu2}=
\left\{
    \begin{aligned}
        &\sum_{p=0}^{P-1} s_{p,R},\ U\left(CLDP\_S_{P,R}^{ri}\right)\leq2\\
        &P+1,\quad\ \ \mbox{Otherwise}
    \end{aligned},
\right.}
\end{equation}
where function $U\left(\cdot\right)$ counts the frequency of bitwise transitions and superscript ``$riu2$'' represents the pattern involving rotation invariance and uniform mapping. When $P=8$, the number of patterns, with regarded as ``features dimension'', has been reduced from $36$ to $10$~\cite{ojala2002multiresolution}.

%
%
\vspace{-0.15in}
\subsubsection{Circular Magnitude Pattern}
\label{sssec:circular magnitude pattern}
Since magnitude information $m_{p,R}=|d_{p,R}|$ of the local difference is complementary to sign information $s_{p,R}$, we obtain the magnitude component of CLDP by applying a thresholding function $t(m_{p,R},c_{m,R})$~\cite{guo2010completed}, where $c_{m,R}$ is the mean magnitude values of local differences of all pixels in the entire image. 
Similar to the coding strategy in circular sign patterns, we define circular magnitude pattern $CLDP\_M_{P,R}^{riu2}$ as follows:
\begin{equation}
\label{eq:CLDP_M_riu2}
\Scale[0.83]{
CLDP\_M_{P,R}^{riu2}
=\left\{
\begin{aligned}
&\sum_{p=0}^{P-1}t(m_{p,R},c_{m,R}), U\left(CLDP\_M_{P,R}^{ri}\right)\leq2\\
&P+1,\qquad\qquad\quad\ \mbox{Otherwise}
\end{aligned}.
\right.}\\
\end{equation}
\subsubsection{Directional Derivative Pattern}
\label{sssec:directional derivative pattern}
To characterize texture smoothness along each direction, we propose a new component, the directional derivative of local difference. Different from CLBP, which involves only one circle, we define two neighboring circles with radii $R$ and $R-1$, respectively, as Fig.~\ref{fig:sampleblock} shows. On the basis of
sign components $s_{p,R}$ and $s_{p,R-1}$ from two neighboring circles, we define directional derivative pattern $CLDP\_D$ as follows:
\begin{equation} \label{eq:CLDP_D}
CLDP\_D_{P,R}=\sum_{p=0}^{P-1}\left(s_{p,R} \oplus s_{p,R-1}\right)\cdot2^p,
\end{equation}
\\
where operator $\oplus$ represents a bitwise exclusive OR (XOR) operation between the sign components of two neighboring circles in the same direction. In the outcome of $s_{p,R} \oplus s_{p,R-1}$, ``1'' means two local differences in one direction have different sign components, and it is likely that the intensity values of the two neighboring pixels vary remarkably. In contrast, ``0'' means the two local differences have the same sign component, which represents certain smoothness. To have the consistent coding format with $CLDP\_S_{P,R}^{riu2}$ and $CLDP\_M_{P,R}^{riu2}$, we define $CLDP\_D_{P,R}^{riu2}$ as follows:
\begin{equation}
\label{eq:CLDP_D_riu2}
\Scale[0.81]{
CLDP\_D_{P,R}^{riu2}
=\left\{
    \begin{aligned}
    &\sum_{p=0}^{P-1}\left(s_{p,R} \oplus s_{p,R-1}\right),U\left(CLDP\_D_{P,R}^{ri}\right)\leq 2\\
    &P+1,\qquad\qquad\qquad\ \mbox{Otherwise}
    \end{aligned}.
    \right.}
\end{equation}
Since we use ``$riu2$'' patterns in all the following sections, for simplification, we denote $CLDP\_S_{P,R}^{riu2}$, $CLDP\_M_{P,R}^{riu2}$, and $CLDP\_D_{P,R}^{riu2}$ as $CLDP\_S$, $CLDP\_M$, and $CLDP\_D$, respectively.
\subsubsection{Global Sign Pattern}
\label{sssec:global sign pattern}
Since all local patterns are extracted based on the local difference, the intensity value of center pixel, $g_c$, which reflects the global information, has been removed. To involve global features, we utilize a binary bit for each center pixel as follows~\cite{guo2010completed}:
\begin{equation} \label{eq:CLDP_C}
CLDP\_C=t(g_c,c_I),
\end{equation}
where $c_I$ represents the mean intensity value of the whole image.\\

\vspace{-0.15in}
\subsection{CLDP Histogram Generation and Classification}
\label{ssec:feature fusion}
\subsubsection{CLDP Histogram Generation}
\label{sssec:histogramgeneration}

 As we discussed in previous sections, $CLDP\_S$, $CLDP\_M$, and $CLDP\_D$ reflect local texture features, and $CLDP\_C$ encodes global information. To combine these four types of patterns, we can utilize three histogram-based schemes: the concatenated, joint, and hybrid histograms, which are similar to the combination schemes in CLBP. In the concatenated histogram, we calculate the histograms of ``$riu2$'' local patterns separately and concatenate their histograms. Such an operation can be denoted as ``$\_$''. For example, when we combine $CLDP\_S$ and $CLDP\_D$ into a concatenated histogram, we denote this combination strategy as $CLDP\_S\_D$. In contrast, the joint histogram first concatenates CLDP codes and then calculates the corresponding histogram. From another perspective, this scheme can be understood as the conversion from a joint multi-dimensional histogram to a 1-D histogram. By defining this operation as ``$/$'', we denote the joint histogram of $CLDP\_M$ and $CLDP\_C$ as $CLDP\_M/C$. The hybrid scheme contains the combination of both the joint and concatenated histograms. For example, $CLDP\_S\_D\_M/C$ contains the concatenated histogram of $CLDP\_S$ and $CLDP\_D$ and the joint histogram of $CLDP\_M$ and $CLDP\_C$. Then, the concatenated histogram of $CLDP\_S\_D$ and $CLDP\_M/C$ forms $CLDP\_S\_D\_M/C$. These three combination schemes of CLDP codes will be used in the Section~\ref{sec:experimental results}.
\vspace{-0.15in}
\subsubsection{Classification}
\label{sssec:classification}

On the basis of the CLDP histograms of texture images, we can implement rotation invariant texture classification. To measure the similarity of two CLDP histograms $T$ and $M$ for test image $I_T$ and model image $I_M$, we use the nearest neighborhood classifier with the chi-square distance~\cite{ojala2002multiresolution} as follows:
\begin{equation}
\label{eq:classifier}
\Scale[1]{
    \begin{aligned}
    D(T,M)=\sum_{n=1}^N\frac{(T_n-M_n)^2}{T_n+M_n}
    \end{aligned},
    }
\end{equation}
where $N$ is the number of bins and $T_n$ and $M_n$ are the values of $T$ and $M$ at the $n$-th bin, respectively. The test image $I_T$ is assigned to the class of $I_M$ that corresponds to the minimal chi-square distance.

\vspace{-0.15in}
\section{Experimental Results}
\label{sec:experimental results}
To evaluate the performance of the proposed method, we focus on the Outex database, which contains 24 classes of texture images with various rotations and illuminations. From the Outex database, we select test suits TC10 and TC12, in which the 24 classes of texture images are captured under 27 conditions, including three types of illumination conditions (``inca", ``t184", and ``horizon") and nine rotation angles ($0^{\circ}$, $15^{\circ}$, $30^{\circ}$, $45^{\circ}$, $60^{\circ}$, $75^{\circ}$, and $90^{\circ}$). Under each situation, every class contains twenty non-overlapping samples. The details of the experiment setup are listed as follows:
\begin{enumerate}[(1)]
    \item For test suit TC10, all samples are captured under the illumination ``inca" and divided into training and testing parts for the evaluation purpose. In each class, samples with rotation angle $0^{\circ}$ are used for classifier training and samples with the remaining eight rotation angles are used for testing. Therefore, the total number of training samples is $24\times20=480$, and the total number of testing samples is $24\times20\times8=3840$.
    \item For test suit TC12, samples are collected under two different illuminations, ``t184" and ``horizon", respectively. Since the training set of TC12 is the same to that of TC10, samples with all rotation angels can be used for testing. Therefore, the total number of testing samples under each illumination is $24\times20\times9=4320$.
\end{enumerate}
\begin{table}[t]

\begin{center}
\caption{Average classification accuracy ($\%$) of CLBP and CLDP on TC10 (``inca'') and TC12 (``t184'' and ``horizon'') using different combination schemes}

\resizebox{0.5\textwidth}{!}{

\begin{tabular}{|c|c|c|c|c|c|c|}

\hline
 Classification  & \multicolumn{6}{c|}{$(P, R)$}  \\
 \cline{2-7}
 Accuracy ($\%$) & $(8, 2)$ & $(8, 3)$ & $(16, 2)$ & $(16, 3)$ & $(24, 2)$ & $(24, 3)$\\
\hline
\hline
$CLBP\_S$ & 77.67 & 80.94 & 82.40 & 85.89 &84.07 & 87.04 \\
\hline
$CLDP\_S/D$ &86.01 & 90.69&89.44 & 92.03& 90.18& 92.55 \\
\hline
$\Delta$ (Accuracy)  & 8.34 & 9.75 & 7.04 & 6.14 & 6.11 & 5.51\\
\hline
\hline
$CLBP\_M$ & 75.45 & 79.32 & 79.96 & 84.35 & 80.35 & 85.12 \\
\hline
$CLDP\_M/D$ & 81.81 & 84.46& 84.88 & 88.56 & 84.80 & 89.24\\
\hline
$\Delta$ (Accuracy)  & 6.36 & 5.14 & 4.92 & 4.21 & 4.45 & 4.12\\
\hline
\hline
$CLBP\_M/C$ & 88.03 & 88.37 & 91.79 & 92.36 & 91.47 & 93.15 \\
\hline
$CLDP\_M/D/C$  & 91.48 & 91.42 & 92.83 & 94.39 &92.37& 94.57 \\
\hline
$\Delta$ (Accuracy) & 3.45 & 3.05 & 1.04 & 2.03 & 0.90 & 1.42\\
\hline
\hline
$CLBP\_S\_M/C$ &92.11  & 92.31 & 93.41 & 94.52& 93.51 & 94.94 \\
\hline
$CLDP\_S\_D\_M/C$ & 92.40 &93.87 & 93.68 &95.24 &93.69 &95.35\\
\hline
$\Delta$ (Accuracy)  & 0.29 & 1.56 & 0.27 & 0.72 & 0.18 & 0.41\\
\hline
\hline
$CLBP\_S/M$ & 92.66 & 94.54& 93.24& 95.00& 93.40& 95.40 \\
\hline
$CLDP\_S/M/D$ & 94.87  & 96.43&95.07 &96.26 & 93.67 & 95.59 \\
\hline
$\Delta$ (Accuracy)  & 2.21 & 1.89 & 1.83 & 1.26 & 0.27 & 0.10\\
\hline
\hline
$CLBP\_S/M/C$  & 95.41 &96.08 &95.44 & 96.16 & 95.19 & \textbf{96.28}\\
\hline
$CLDP\_S/M/D/C$ & 96.29 & \textbf{97.14} &96.25 & 96.45 & 94.94 & 95.97 \\
\hline
$\Delta$ (Accuracy)  & 0.88 & 1.06 & 0.81 & 0.29 & -0.25 & -0.31\\
\hline
\end{tabular}
}
\label{tab:comparison of clbp vs cldp}
\end{center}
\end{table}

\vspace{-0.05in}

Because of the efficiency and robustness of CLBP, we choose it as an important benchmark and compare its classification accuracy with that of the proposed method in different combination schemes as Table~\ref{tab:comparison of clbp vs cldp} shows. Each row in Table~\ref{tab:comparison of clbp vs cldp} represents the classification accuracy under a certain scheme with six types of $(P, R)$ changing from $(8, 2)$ to $(24, 3)$. For each scheme of CLBP, we create a corresponding scheme that adds our newly proposed $CLDP\_D$ using joint or concatenated histograms. Each entry in Table~\ref{tab:comparison of clbp vs cldp} represents the average classification accuracy over three test suits mentioned above. We notice that CLDP improves the classification accuracy of CLBP counterparts for all $(P,R)$ selections except for the last two schemes at the last two columns. Along with the changing of $(P, R)$, CLDP with fewer sampling points has more accuracy improvement. This implicates that adding the directional derivative pattern is the most helpful when the number of sampling points is smaller.
In addition, more complicated combination schemes correspond to less accuracy improvement.
The superior performance of CLDP proves that  $CLDP\_D$ describes the variations of texture along radial directions and provides complementary information to $CLDP\_S$, $CLDP\_M$, and $CLDP\_C$. Under scheme $CLDP\_S/M/D/C$ with $(P,R)=(8,3)$, CLDP achieves its highest accuracy rate, $97.14\%$, which is $0.86\%$ higher than that of CLBP, $96.28\%$, in scheme $CLBP\_S/M/C$ with $(P,R)=(24,3)$. In addition, the feature dimension of $CLDP\_S/M/D/C$ with $(P,R)=(8,3)$, $2\times 10^3=2000$, is only one third of that of $CLBP\_S/M/C$ with $(P,R)=(24,3)$, $26^2\times10=6760$. Therefore, CLDP with lower $(P,R)$ has better performance in both accuracy and efficiency.

In addition to CLBP, we compare CLDP with other texture descriptors and list the corresponding classification accuracy on three test suits in Table~\ref{tab:classificatoin accuracy using popular texture descriptors}. Since the training and testing data of TC12 have different illumination, we average the classification accuracy of the only two test suits in TC12 without involving that of TC10 for fair comparison. In Table~\ref{tab:classificatoin accuracy using popular texture descriptors}, the upper and lower parts correspond to uni- and multi-scale operators, respectively. We notice that, the proposed method, being a uni-scale operator, has the best performance when compared with other uni-scale state of the art. In addition, the proposed method can also achieve comparable accuracy with multi-scale operators, which increase accuracy by sacrificing computational efficiency.

\begin{table}[t]

\begin{center}
\caption{Classification accuracy ($\%$) of state-of-the-art texture descriptors on TC10 (``inca'') and TC12 (``t184'' and ``horizon''). Accuracies are as originally reported, expect those for PRICoLBP$_g$ which are taken from the work by Liu et al.~\cite{liu2015median}.}

\resizebox{0.5\textwidth}{!}{

\begin{tabular}{|c|c|c|c|c|}

\hline
Classification & \multirow{2}{*}{\centering TC10  }  &\multicolumn{3}{c|}{TC12} \\
\cline{3-5}
  Accuracy (\%) &   & t184 & horizon  & Average \\
\hline
\hline
 CLBP~\cite{guo2010completed} & 99.30 & 95.32 & 94.54  & 94.93\\
\hline
DLBP + NGF ~\cite{liao2009dominant}& 99.10 & 93.20 & 90.40  & 91.80\\
\hline
PRICoLBP$_g$~\cite{qi2014pairwise} & 94.48 & 92.57 & 92.50  & 92.54\\
\hline
LDDP~\cite{guo2012local} & 97.89 & 95.30 & 93.40  & 94.35\\
\hline
CLBC~\cite{zhao2012completed} & 98.80 & 94.00 & 94.07 & 94.04\\
\hline
DNS + LBP$_{(24,3)}~\cite{khellah2011texture}$ & 99.27 & 94.40 & 93.85 & 93.63\\
\hline
PLBP\_C~\cite{shadkam2012local} & 98.95 & 95.32 & 94.95 & 95.14\\
\hline
NTLBP~\cite{fathi2012noise} & 99.24 & 96.18 & 94.28 & 95.23\\

\hline
\hline
CLBP\_S/M/C$_{(8,1)+(16,2)+(24,3)}$ & \multirow{2}{*}{\centering 99.14} & \multirow{2}{*}{\centering 95.18} & \multirow{2}{*}{\centering 95.55}  & \multirow{2}{*}{\centering 95.37}\\
(Multi-scale)~\cite{guo2010completed} & & & &\\
\hline
LDDP\_1/2$_{(8,1)+(16,2)+(24,3)}$ & \multirow{2}{*}{\centering 98.64} & \multirow{2}{*}{\centering 95.9} & \multirow{2}{*}{\centering 94.16}  & \multirow{2}{*}{\centering 95.08}\\
(Multi-scale)~\cite{guo2012local} & & & &\\
\hline
CLBC\_S/M/C$_{(8,1)+(16,2)+(24,3)}$ & \multirow{2}{*}{\centering 99.38} & \multirow{2}{*}{\centering 94.98} & \multirow{2}{*}{\centering 95.51}  & \multirow{2}{*}{\centering 95.25}\\
(Multi-scale)~\cite{zhao2012completed} & & & &\\
\hline
MSJ-LBP (Multi-scale)~\cite{qi2013multi} & 96.53 & 94.95 & 96.34  & 95.65\\
\hline
pi-LBP (Multi-scale)~\cite{lin2015multi} & 99.17 & 95.72 & 94.54 & 95.13\\
\hline
pi-LBP/C (Multi-scale)~\cite{lin2015multi} & 98.96 & 97.36 & 97.11 & 97.24\\
\hline
\hline


CLDP (Proposed) & 99.32 & 96.55 & 95.63  & 96.09\\
\hline
\end{tabular}
}
\label{tab:classificatoin accuracy using popular texture descriptors}
\end{center}
\end{table}

\section{Conclusions}
\label{sec:conclusions}
We proposed a new texture descriptor, CLDP, on the basis of CLBP. CLDP involves four patterns, $CLDP\_S$, $CLDP\_M$, $CLDP\_D$, and $CLDP\_C$, in which the last pattern, as a complement to the other three, reflects the smoothness of local texture. We combine these patterns using histogram-based schemes for high accuracy on texture classification. Experimental results showed that the proposed CLDP has the best performance in contrast to other state-of-the-art uni-scale descriptors. In our future research, we will extend CLDP to a multi-scale version.  By involving more local information, the multi-scale-based CLDP may have better performance on texture classification. Furthermore, we will also explore ways to reduce feature dimensions.

%

\clearpage

%

\bibliographystyle{IEEEbib}
\bibliography{refs}

\end{document}